\documentclass[aps, 10pt, pr,notitlepage, twocolumn, superscriptaddress,nofootinbib]{revtex4-1}

\usepackage{amsmath}
\usepackage{amssymb}
\usepackage{amsfonts}
\usepackage[utf8]{inputenc}
\usepackage[T1]{fontenc}
\usepackage{mathrsfs}
\usepackage{anyfontsize}

\usepackage[usenames,dvipsnames]{xcolor}

\usepackage{newtxtext}
\usepackage{newtxmath}
\usepackage{tensor}
\usepackage{bm}

\usepackage{graphicx}
\usepackage{epsfig}
\usepackage{epstopdf}

\usepackage{natbib}

\usepackage[normalem]{ulem}

\def\jnl@style{\it}
\def\aaref@jnl#1{{\jnl@style#1}}

\def\aaref@jnl#1{{\jnl@style#1}}

\def\aj{\aaref@jnl{AJ}}                   
\def\apj{\aaref@jnl{ApJ}}                 
\def\apjl{\aaref@jnl{ApJ}}                
\def\apjs{\aaref@jnl{ApJS}}               
\def\apss{\aaref@jnl{Ap\&SS}}             
\def\aap{\aaref@jnl{A\&A}}                
\def\aapr{\aaref@jnl{A\&A~Rev.}}          
\def\aaps{\aaref@jnl{A\&AS}}              
\def\mnras{\aaref@jnl{Mon.~Not.~Roy.~Astron.~Soc.}}             
\def\prd{\aaref@jnl{Phys.~Rev.~D}}        
\def\prc{\aaref@jnl{Phys.~Rev.~C}}  
\def\prl{\aaref@jnl{Phys.~Rev.~Lett.}}    
\def\qjras{\aaref@jnl{QJRAS}}             
\def\skytel{\aaref@jnl{S\&T}}             
\def\ssr{\aaref@jnl{Space~Sci.~Rev.}}     
\def\zap{\aaref@jnl{ZAp}}                 
\def\nat{\aaref@jnl{Nature}}              
\def\aplett{\aaref@jnl{Astrophys.~Lett.}} 
\def\apspr{\aaref@jnl{Astrophys.~Space~Phys.~Res.}} 
\def\physrep{\aaref@jnl{Phys.~Rep.}}      
\def\physscr{\aaref@jnl{Phys.~Scr}}       
\def\commat{\aaref@jnl{Comm.~Math.~Phys.}}              
\def\science{\aaref@jnl{Science}}               
\def\cqg{\aaref@jnl{Classical Quant.~Grav.}}            
\def\jpcs{\aaref@jnl{JPCS}}                                     
\def\ijmpd{\aaref@jnl{Int.~J.~Mod.~Phys.~D}}                    
\def\grg{\aaref@jnl{Gen.~Relat.~Gravit.}}               
\def\rpp{\aaref@jnl{Rep.~Prog.~Phys.}}          
\def\npa{\aaref@jnl{Nucl.~Phys.~A}}        
\def\lrr{\aaref@jnl{Living Rev.~Rel.}}                   
\def\jcap{\aaref@jnl{J.~Cosmology Astropart.~Phys.}}    
\def\rmp{\aaref@jnl{Rev.~Mod.~Phys.}}   

\allowdisplaybreaks[1]

\begin{document}

\title{Non-topological spontaneously scalarized neutron stars in tensor-multi-scalar theories of gravity }

\author{Daniela D. Doneva}
\email{daniela.doneva@uni-tuebingen.de}
\affiliation{Theoretical Astrophysics, Eberhard Karls University of T\"ubingen, T\"ubingen 72076, Germany}
\affiliation{INRNE - Bulgarian Academy of Sciences, 1784  Sofia, Bulgaria}

\author{Stoytcho S. Yazadjiev}
\email{yazad@phys.uni-sofia.bg}
\affiliation{Theoretical Astrophysics, Eberhard Karls University of T\"ubingen, T\"ubingen 72076, Germany}
\affiliation{Department of Theoretical Physics, Faculty of Physics, Sofia University, Sofia 1164, Bulgaria}
\affiliation{Institute of Mathematics and Informatics, 	Bulgarian Academy of Sciences, 	Acad. G. Bonchev St. 8, Sofia 1113, Bulgaria}


\begin{abstract}
 In the present paper we numerically construct  new non-topological, spontaneously scalarized neutron stars in the tensor-multi-scalar theories of gravity whose target space is a three-dimensional maximally symmetric space, namely either $\mathbb{S}^3$, $\mathbb{H}^3$ or $\mathbb{R}^3$, and  in the case of a nontrivial map  $\varphi : \text{\it spacetime} \to \text{\it target space}$. The theories of gravity admitting scalarization are characterized by the fact that the field equations always admit the general relativistic solution but for certain ranges of the parameters space it looses stability and nonlinear development of a scalar field is observed. Thus, in order to determine the values of the parameters where such scalarization is possible we studied the stability of the general relativistic solution within the framework of the considered tensor-multi-scalar theories. Based on these results we could obtain a family of scalarized branches characterized by the number of the scalar field nodes. These branches bifurcate from the general relativistic solution at the points where new unstable modes appear and they are energetically more favorable over the pure Einstein solutions.  Interestingly, in certain parameter ranges we could obtained non-uniqueness within a single branch of scalarized solutions. 
\end{abstract}


\maketitle

\section{Introduction}

The era of gravitational waves (GW) is beginning and a lot of new detections of GW are
 expected  at increasing precision in the next years. The network of gravitational detectors will soon be expanded with new detectors
namely with KAGRA in Japan and with LIGO-India. The expanded gravitational detectors network will be in principle capable to 
probe precisely even the polarization content of GWs. This will be very important for constraining the modified gravitational theories 
which, as a role, predict GWs possessing more than the two polarizations in GR. The sources of GW are usually related 
to the dynamics of the compact objects in the gravitational theories and the GW emission depends strongly on the type, structure and properties 
of the compact objects.  

A particular interest attract theories possessing  scalar degrees of freedom which are perturbatively equivalent to general relativity in the weak field regime and thus always admit the general relativistic solutions, but can lead to large deviations for strong fields through the nonlinear development of a scalar field, that is the so-called spontaneous scalarization. Such scalarization was first observed 
for neutron stars in usual scalar-tensor theories with only one scalar field \cite{Damour2013} and then for black holes \cite{Stefanov2008}--\cite{Cardoso2013}. Interestingly, it was recently shown that the spacetime curvature can also act as a source for the spontaneous scalarization in Gauss-Bonnet gravity \cite{Doneva2017} -- \cite{Doneva:2017duq}

In the present paper we focus on tensor-multi-scalar theories of gravity (TMST). These modified gravitational theories are mathematically self-consistent and can pass all known experimental  and observational tests with a proper choice of the functions and the parameters \cite{Damour_1992,Horbatsch_2015}. In TMST  various compact objects can be constructed \cite{Yazadjiev:2019oul}--\cite{Doneva:2019ltb} where the freedom comes not only from the choice of the conformal factor like the standard scalar-tensor theories, but more importantly from the choice of the target space and the  metric defined on it, as well as from the choice of the map $\varphi : \text{\it spacetime} \to \text{\it target space}$.

The purpose of the present paper is to present new non-topological, spontaneously scalarized neutron stars in the TMST of gravity whose target  space is a 3-dimensional maximally symmetric space, namely $\mathbb{S}^3$, $\mathbb{H}^3$ or $\mathbb{R}^3$ and in the case of a nontrivial map 
$\varphi : \text{\it spacetime} \to \text{\it target space}$. The mathematical construction of the problem is very similar to the topological neutron stars presented in \cite{Doneva:2019ltb}, with the very important difference that we require zero boundary condition for the scalar field at the origin. This leads to zero topological charge of the neutron stars, but the theory admits scalarization that will be our focus. Such  non-topological, spontaneously scalarized neutron stars posses some attractive features which make them interesting and place them among the realistic objects that could exist in Nature. For example they possess zero scalar charge and thus  evades the strong
binary pulsar constraints on dipole scalar radiation -- in other words they behave as the ordinary general relativistic 
neutron stars in the binary systems.

\section{Tensor-multi-scalar theories of gravity}    

In TMST the gravitational interaction is mediated  by the spacetime metric $g_{\mu\nu}$ and $N$ scalar fields $\varphi^{a}$  which take value in a coordinate patch of an N-dimensional Riemannian (target) manifold ${\cal E}_{N}$  with positively definite metric $\gamma_{ab}(\varphi)$ defined on it \cite{Damour_1992,Horbatsch_2015}. The Einstein frame action of the  TMST of gravity is given by 
\begin{eqnarray}\label{Action}
S=&& \frac{1}{16\pi G_{*}}\int d^4\sqrt{-g}\left[R - 2g^{\mu\nu}\gamma_{ab}(\varphi)\nabla_{\mu}\varphi^{a}\nabla_{\nu}\varphi^{b} - 4V(\varphi)\right]  \nonumber \\
&&+ S_{matter}(A^{2}(\varphi) g_{\mu\nu}, \Psi_{matter}),
\end{eqnarray}
where $G_{*}$ is the bare gravitational constant, $\nabla_{\mu}$ and $R$ are the covariant derivative  and the Ricci scalar curvature with respect to  the Einstein frame metric $g_{\mu\nu}$, and $V(\varphi)\ge 0$ is the potential of the scalar fields. In order for the weak equivalence principle to be satisfied the matter fields, denoted collectively by $\Psi_{matter}$, are coupled only to the physical Jordan frame metric ${\tilde g}_{\mu\nu}= A^2(\varphi) g_{\mu\nu}$ where  $A^2(\varphi)$ is the conformal factor relating the Einstein and the Jordan frame metrics, and which, together with $\gamma_{ab}(\varphi)$ and $V(\varphi)$, specifies the TMST. From a more global point of view $\varphi^a$ 
define a map $\varphi : \text{\it spacetime} \to \text{\it target space}$ and the scalar fields kinetic term in the action above is just the pull-back  of the line element of the target space.  

The Einstein frame field equations  corresponding to the action (\ref{Action}) are the following  
\begin{eqnarray}\label{FE}
&&R_{\mu\nu}= 2\gamma_{ab}(\varphi) \nabla_{\mu}\varphi^a\nabla_{\nu}\varphi^b + 2V(\varphi)g_{\mu\nu} + 8\pi G_{*} \left(T_{\mu\nu} - \frac{1}{2}T g_{\mu\nu}\right), \nonumber \\
&&\nabla_{\mu}\nabla^{\mu}\varphi^a = - \gamma^{a}_{\, bc}(\varphi)g^{\mu\nu}\nabla_{\mu}\varphi^b\nabla_{\nu}\varphi^c 
+ \gamma^{ab}(\varphi) \frac{\partial V(\varphi)}{\partial\varphi^{b}}  \\
&&\hskip 1.6cm -  4\pi G_{*}\gamma^{ab}(\varphi)\frac{\partial\ln A(\varphi)}{\partial\varphi^{b}}T, \nonumber
\end{eqnarray}
with $T_{\mu\nu}$ being  the Einstein frame energy-momentum tensor of matter and $\gamma^{a}_{\, bc}(\varphi)$ being 
the Christoffel symbols with respect to the target space metric $\gamma_{ab}(\varphi)$. From the field equations and the contracted Bianchi identities we also find the following conservation law for the Einstein frame energy-momentum tensor
\begin{eqnarray}\label{Bianchi}
\nabla_{\mu}T^{\mu}_{\nu}= \frac{\partial \ln A(\varphi)}{\partial \varphi^{a}}T\nabla_{\nu}\varphi^a .
\end{eqnarray}

The Einstein frame energy-momentum tensor $T_{\mu\nu}$ and the Jordan frame one ${\tilde T}_{\mu\nu}$
are related via the formula $T_{\mu\nu}=A^{2}(\varphi){\tilde T}_{\mu\nu}$. As usual, in the present paper  the matter content of 
the stars will be described as a perfect fluid. In the case of a perfect fluid the relations between the energy density,
pressure and 4-velocity in both frames are given by $\varepsilon=A^{4}(\varphi){\tilde \varepsilon}$, $p=A^{4}(\varphi){\tilde p}$ and 
$u_{\mu}=A^{-1}(\varphi) {\tilde u}_{\mu}$.

We are interested in strictly static (in both the Einstein and the Jordan frame), completely regular, spherically symmetric  and asymptotically flat solutions to the equations of the TMST describing neutron stars.  The spacetime metric can  then be written in  the standard form
\begin{eqnarray}
ds^2= - e^{2\Gamma}dt^2 + e^{2\Lambda}dr^2 + r^2(d\theta^2  + \sin^2\theta d\phi^2)
\end{eqnarray} 
where $\Gamma$ and $\Lambda$ depend on the radial coordinate $r$ only.

In the present paper, as we have already mentioned, we shall consider TMST whose target space manifold is a 3- dimensional symmetric space, namely 
$\mathbb{S}^3$, $\mathbb{H}^3$ or $\mathbb{R}^3$
  with the metric 
\begin{eqnarray}
\gamma_{ab}d\varphi^a d\varphi^b= a^2\left[d\chi^2 + H^2(\chi)(d\varTheta^2 + \sin^2\varTheta d\Phi^2) \right],
\end{eqnarray}
where $a>0$ is a constant  and $\varTheta$ and $\Phi$ are the standard angular coordinates on the 2-dimensional sphere $\mathbb{S}^2$.
The function $H(\chi)$ is given by $H(\chi)=\sin\chi$ for the spherical geometry, $H(\chi)=\sinh\chi$ for the hyperbolic geometry and
$H(\chi)=\chi$ in the case of the flat geometry. The parameter $a$ is related to the curvature $\kappa$ of  $\mathbb{S}^3$ and $\mathbb{H}^3$.  
We have $\kappa=1/a^2$ for spherical  and $\kappa=-1/a^2$ for hyperbolic geometry.   Our choice of  the target spaces is motivated by the fact that the round $\mathbb{S}^3$, $\mathbb{H}^3$ or $\mathbb{R}^3$ are among the simplest target spaces admitting spherically symmetric neutron star solutions for the ansatz define below. In addition we shall consider theories for which the coupling function $A(\varphi)$ and the potential $V(\varphi)$ depend on $\chi$ only. This allows the equations for $\Theta$ and $\Phi$ to separate form the main system and  guarantees that the spacetime metric will be spherically symmetric in both the Einstein and the Jordan frame for the ansatz defined below. 

Usually the people choose the simplest map $\varphi : \text{\it spacetime} \to \text{\it target space}$ for which all the scalar fields depend 
on the radial coordinate $r$ only.  Here we choose a nontrivial map $\varphi : \text{\it spacetime} \to \text{\it target space}$ defined as follows.
We  assume that the field $\chi$ depends on the radial coordinate $r$, i.e. $\chi=\chi(r)$, and the fields $\varTheta$ and $\Phi$ are independent from $r$ and are given by $\Theta=\theta$ and $\Phi=\phi$ \cite{Doneva:2019ltb}. Our ansatz is  compatible with the spherical symmetry and one can  check that the equations for  $\varTheta$ and $\Phi$ are satisfied.

With this ansatz in mind for the dimensionally reduced field equations we have:

\begin{widetext}
\begin{eqnarray} 
&&\frac{2}{r}e^{-2\Lambda} \Lambda^{\prime} + \frac{1}{r^2}\left(1-e^{-2\Lambda}\right)=8\pi G_{*} A^{4}(\chi) {\tilde \varepsilon}
+a^2 \left(e^{-2\Lambda} {\chi^{\prime}}^2 + 2 \frac{H^2(\chi)}{r^2}\right)  + 2V(\chi), \label{DRE} \\ 
&&\frac{2}{r}e^{-2\Lambda} \Gamma^{\prime} - \frac{1}{r^2}\left(1-e^{-2\Lambda}\right)=8\pi G_{*} A^{4}(\chi) {\tilde p}
+a^2 \left(e^{-2\Lambda} {\chi^{\prime}}^2 - 2 \frac{H^2(\chi)}{r^2}\right)  - 2V(\chi), \\
&&\chi^{\prime\prime} + \left(\Gamma^\prime - \Lambda^\prime + \frac{2}{r}\right)\chi^{\prime}= \left[\frac{2}{r^2}H(\chi)\frac{\partial H(\chi)}{\partial\chi}  + \frac{1}{a^2} \frac{\partial V(\chi)}{\partial\chi} + \frac{4\pi G_{*}}{a^2} A^4(\chi)\frac{\partial \ln A(\chi)}{\partial\chi}(\tilde{\varepsilon} - 3{\tilde p})\right]e^{2\Lambda},\\
&& {\tilde p}^\prime = - (\tilde{\varepsilon} + {\tilde p}) \left[\Gamma^\prime + \frac{\partial \ln A(\chi)}{\partial\chi} \chi^\prime \right], \label{HSE}
\end{eqnarray} 
\end{widetext}
where the prime denotes differentiation with respect to $r$. In what follows we shall restrict ourself to the case with $V(\chi)=0$.  

Asymptotic flatness requires $\Gamma(\infty)=\Lambda(\infty)=0$ and $\chi(\infty)=0$. Regularity at the center requires $\Lambda(0)=0$ and $\chi(0)=0$. When the target space is $\mathbb{S}^3$ the regularity condition in the center for the scalar field $\chi$ is more general \cite{Doneva:2019ltb},
namely $\chi(0)=n\pi$ with $n\in \mathbb{Z}$. However the focus of this paper are the non-topological scalarized neutron stars and that is why we put $n=0$.

The above system of equations (\ref{DRE})--(\ref{HSE}) supplemented with the equation of state of the baryonic matter ${\tilde p}={\tilde p}({\tilde \varepsilon})$, with the above mentioned asymptotic and regularity conditions as  well as with a specified central energy density ${\tilde \varepsilon}_c$, describes the structure of the neutron stars in the TMST under consideration.  

The mass $M$ of the neutron star is defined as the ADM mass in the Einstein frame and can be extracted from the asymptotic behavior of the metric functions $\Gamma$ and $\Lambda$, namely 
\begin{eqnarray}
\Gamma\approx -\frac{M}{r} + O(1/r^2),\,\,\,  \Lambda \approx \frac{M}{r} + O(1/r^2).
\end{eqnarray}
The asymptotic behavior of the scalar field $\chi$  can be derived from the linearized  equation for $\chi$ outside the star and  is given by    
\begin{eqnarray}
\chi\approx \frac{const}{r^2} + O(1/r^3).
\end{eqnarray}
This means that the scalar charge $D_{\chi}$ of the neutron star associated with $\chi$ is zero. The fact that $\chi$ drops like $1/r^2$  leads also to the conclusion that  the ADM masses in both frames are the same.

\section{Scalarized neutron stars}

In the present paper we shall focus on TMST with coupling function $A(\chi)$ satisfying the conditions

\begin{eqnarray}\label{CA}
\frac{\partial A}{\partial \chi}(0)=0, \,\,\,  \frac{\partial^2 A}{\partial \chi^2}(0)\ne 0.
\end{eqnarray}

Taking into account the boundary and the regularity conditions discussed above as well as the first of conditions (\ref{CA}), it is not difficult one to see that the general relativistic neutron star solutions are also solutions to  our system of equations (\ref{DRE})--(\ref{HSE}) but with  a trivial scalar field $\chi=0$. As in the case of usual scalar-tensor theories with only one scalar field one should expect that these general relativistic solutions are unstable  in the framework of the TMST under consideration for certain ranges of the neutron star central energy densities. In order to show this we consider the perturbations of the 
general relativistic neutron stars  within the framework of the described class of TMST.  Of course, the equations governing the perturbations of the metric  are decoupled from the equations governing the perturbations of the scalar fields. Even more, the perturbations of the scalar fields $\Theta$ and $\Phi$ are decoupled from the perturbations of the scalar field $\chi$ and are in fact ordinary wave equations on the general relativistic background.  Therefore, in order to study the possible instability of the general relativistic solutions we have to concentrate on the equation for the perturbations of $\chi$. Since the focus of this paper are spherically symmetric solutions we present the equation governing 
the spherically symmetric perturbations of $\chi$ in the form $\delta\chi(r)=\frac{\delta u(r)}{r} e^{i\omega t}$, namely 

\begin{eqnarray}\label{eq:PertEq}
e^{\Gamma_0 -\Lambda_0}\frac{d}{dr}\left(e^{\Gamma_0 -\Lambda_0}\frac{d\delta u(r)}{dr}\right) + \left(\omega^2 - U(r)\right)\delta u(r)=0
\end{eqnarray}
where the potential $U(r)$ is given by 

\begin{eqnarray}
U(r)= e^{2\Gamma_0}\left[\frac{2}{r^2} + 4\pi G_{*}(\tilde{\epsilon}_0 - 3{\tilde p}_0)\frac{\tilde \beta}{a^2} \right. \nonumber \\
\left. + \frac{1-e^{-2\Lambda_0}}{r^2}  -
4\pi G_{*}(\tilde{\epsilon}_0 - {\tilde p}_0) \right].
\end{eqnarray}
Here all quantities with subscript "0" refer to the general relativistic background and ${\tilde \beta}=\frac{\partial^2\ln A}{\partial\chi^2}(0)$. Let us note that the effective potential is the same for all the functions $H(\chi)$.

\section{Results}
The neutron star solutions will be calculated using the realistic nuclear matter APR4 equation of state \cite{APR4} and we employed its piecewise polytropic approximation \cite{PPA_Read}. We concentrate on two coupling functions $A(\chi)=e^{\frac{1}{2}\beta \chi^2}$ and $A(\chi)=e^{\beta \sin^2{\chi}}$, which have already been employed in the calculation of the topological neutron stars where $n>0$ \cite{Doneva:2019ltb} and thus can be used for comparison.

The numerical solutions describing scalarized neutron stars were obtained after solving the system of reduced field equations \eqref{DRE}--\eqref{HSE} together with the appropriate boundary conditions using a shooting method. Similar to \cite{Doneva:2019ltb}, the shooting parameters are the derivative of the scalar field and the value of the metric function $\Gamma$ at the origin. They are determined from the requirement that $\chi$ and $\Gamma$ vanish at infinity. 

The spectrum of solutions in this class of alternative theories has the following features. The general relativistic solution with zero scalar field is always a solution of the field equations \eqref{DRE}--\eqref{HSE}. In addition, in a certain range of the parameter space, nontrivial scalar field can develop and the scalarized solutions coexist with the non-scalarized ones. Thus, we have nonuniqueness and the problem is very similar from a mathematical point of view to the scalarization in the standard scalar-tensor theories with one scalar field \cite{Damour_1992}.  

In order to understand the spectrum of solutions and to check the stability of the general relativistic neutron stars with zero scalar field, one has to  solve the perturbation equation \eqref{eq:PertEq}. The first point where $\omega=0$ signals the appearance of the first unstable mode of the GR solutions but of course more unstable modes can exist for larger central energy densities and they are classified by the number of the $\delta \chi$ nodes. It can be easily demonstrated that close to the points where $\omega=0$ the solution for $\delta \chi$ is a solution of the reduced field equations as well. Therefore, new solutions with nontrivial scalar field bifurcate from the GR sequences at the central energy densities for which $\omega=0$. From a practical point of view, one searches for unstable modes with $\omega^2<0$ within the GR sequence of non-scalarized solutions and a change of stability is observed at the central energy density where such modes can no longer be obtained. This is numerically much easier task since the boundary conditions for the perturbation of the scalar field $\delta \chi$ reduce to zero at the origin and at infinity. Thus, we have a standard Sturm-Liouville problem that is not difficult to solve. 

\begin{figure}
	\includegraphics[width=0.25\textwidth]{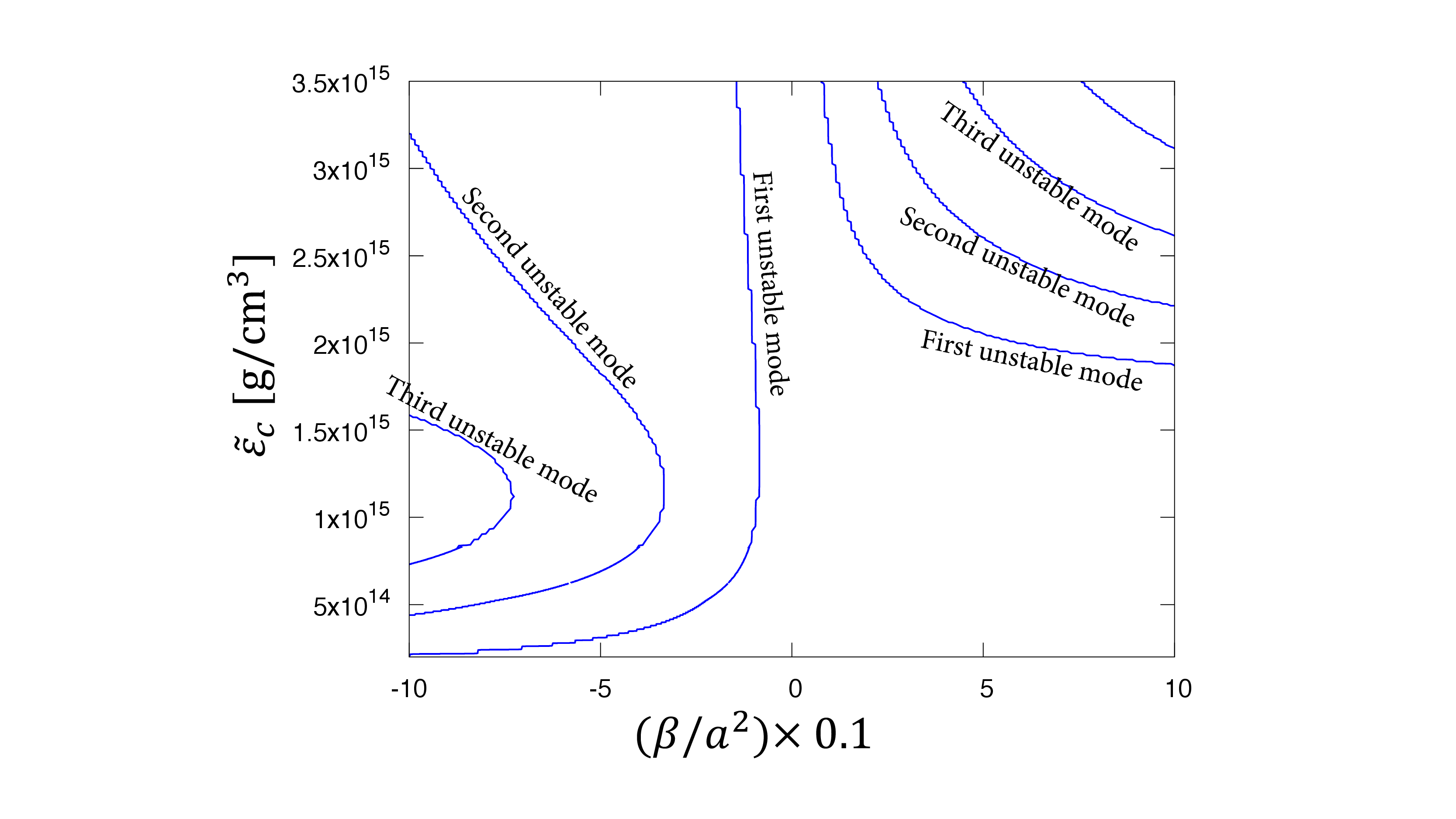}
	\caption{Lines of instability corresponding to zero frequency modes, i.e. $\omega=0$, where on the $x$ axis we have depicted the ratio of the parameters $\beta/a^2$, and on the $y$-axis -- the central energy density ${\tilde \varepsilon}_c$. An unstable mode exists only on the left (right) of the corresponding instability line for $\beta<0$ ($\beta>0$).}
	\label{fig:Bifurcations}
\end{figure}

As one can see, the parameters coming from the TMST enter in the perturbation equation \eqref{eq:PertEq} only through the combination ${\tilde \beta}/a^2$. Therefore, the value of this ratio determines uniquely the central energy densities where change of stability (and bifurcation of scalarized solutions) is observed. Fig. \ref{fig:Bifurcations} represents a two-dimensional plot of the instability lines connecting models having an $\omega=0$ mode for conformal factor $A(\chi)=e^{\beta \sin^2{\chi}}$. Clearly, more than one unstable mode can exist and thus more than one instability line is present. These unstable modes can be labeled by the number of zeros of the scalar field perturbation where the first fundamental mode has no nodes, the second one has one node and so on. Models located to the left (right) on an instability line for $\beta<0$ ($\beta>0$)  posses an unstable mode with the corresponding number of nodes.   The figure is independent of $H$ since it does not enter explicitly in the perturbation equation \eqref{eq:PertEq}. On the $x$ and $y$ axis we have plotted the ratio $\beta/a^2$  and the central energy density  ${\tilde \varepsilon}_c$ respectively. For this conformal factor ${\tilde \beta}=2\beta$.

Let us first focus on the ${\beta}<0$ case. As Fig. \ref{fig:Bifurcations} shows, there are no unstable modes above certain  ${ \beta}/a^2$ and the GR solutions are stable. If we decrease  ${ \beta}/a^2$, the first instability line  appears and for a fixed $\beta/a^2$ there are two central energy densities for which $\omega=0$. Clearly this will translate in two bifurcation points when solving the field equations -- one at small  ${\tilde \varepsilon_c}$ where a scalarized branch of neutron stars appears and the general relativistic solutions loose their stability, and a second one at large  ${\tilde \varepsilon_c}$ where the scalarized branch merges again with the general relativistic solution. While the first bifurcation point moves to lower and lower central energy densities with the decrease of  ${ \beta}/a^2$, the second one moves to large ${\tilde \varepsilon_c}$ and eventually reaches  central energy densities that are considered unphysical.  The first instability line appearing at larger ${ \beta}/a^2$ corresponds to modes having no zeros of the scalar field perturbation, the second instability line corresponds to modes having one zero and so on. Thus, theoretically, an infinity number of instability lines exist appearing when  ${ \beta}/a^2$ is further and further decreased. 

The case with ${\beta}>0$ is qualitatively  similar -- below certain $\beta/a^2$ the GR neutron stars are stable but when $\beta/a^2$  is increased instability lines appear that are associated with the appearance of new unstable modes of the GR solutions. These unstable modes can be labeled by the number of zeros of the scalar field perturbation.  The main difference with the picture for negative ${ \beta}/a^2$ is that for every instability line and for a fixed $\beta/a^2$ we could find only one central energy density where $\omega=0$, if we restrict ourselves to physically reasonable $\tilde{\varepsilon}_c$. Therefore, within a reasonable range of central energy densities, only one   bifurcation point will be present when solving the field equations.  We have produced similar pictures for other coupling functions and EOS, and the results are qualitative the same. 

\begin{figure}
	\includegraphics[width=0.40\textwidth]{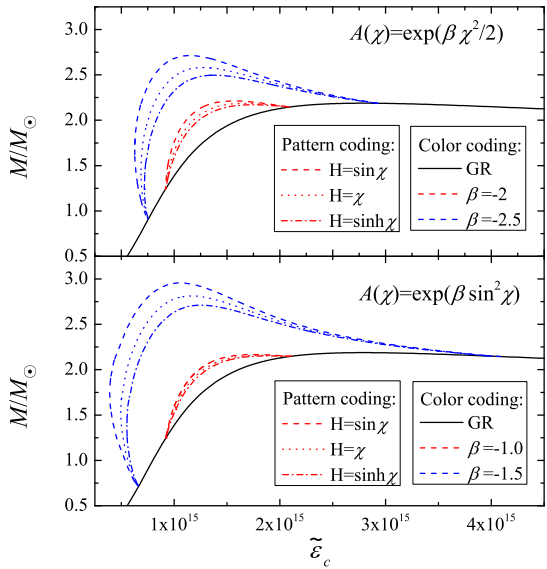}
	\caption{The mass as a function of the central energy density for the first branch of scalarized neutron stars possessing scalar field with no nodes. Solutions for different  choices of $A(\chi)$, $H(\chi)$ and $\beta$ are shown where we have fixed $a^2=0.1$.}
	\label{fig:M_rhoc}
\end{figure}

\begin{figure}
	\includegraphics[width=0.40\textwidth]{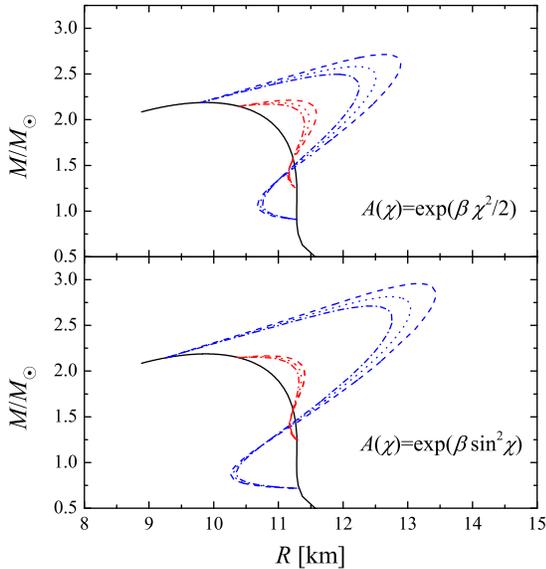}
	\caption{The mass as a function of the neutron star radius for the same solutions as on Fig. \ref{fig:M_R}.}
	\label{fig:M_R}
\end{figure}

Fig. \ref{fig:Bifurcations}  is very useful for finding the solutions of the nonlinear problem defined by the reduced field equations \eqref{DRE}--\eqref{HSE}. The reason is that in the immediate vicinity of a bifurcation point the solution for $\delta \chi$ is also a solution of the field equations \eqref{DRE}--\eqref{HSE} which simplifies the construction of scalarized neutron stars considerably. The first branch of solutions having scalar field with no nodes and corresponding to the first line of instability is plotted in Figs. \ref{fig:M_rhoc} and \ref{fig:M_R} for different combinations of $A(\chi)$, $H(\chi)$ and $\beta$. In these plots we have fixed $a^2=0.1$. As one can see the qualitative picture for different $A(\chi)$ and $H(\chi)$ is very similar. For larger $\beta$ the scalarized branch is characterized by a neutron star mass that first monotonically increases with the increase of ${\tilde \varepsilon_c}$ and after reaching a maximum it starts to decrease. The scalarized neutron stars are uniquely determined by their central energy density. For smaller $\beta$ a region in the low ${\tilde \varepsilon_c}$ domain exists where we have non-uniqueness of the scalarized solutions -- after the first bifurcation point the branch moves towards smaller ${\tilde \varepsilon_c}$ while the mass increases and after reaching a minimum of ${\tilde \varepsilon_c}$, the central energy density starts to increase until the second bifurcation point is reached. As a matter of fact in this region of central energy densities the general relativistic solutions is stable as we have seen from the analysis of the perturbation equation \eqref{eq:PertEq}. Therefore, we have three neutron star solutions (two scalarized ones and one general relativistic) for a fixed ${\tilde \varepsilon_c}$ which makes the stability study of the scalarized neutron stars very important.  The minimum ${\tilde \varepsilon_c}$ where scalarized neutron stars exist depends strongly on $\beta$ and it quickly moves towards smaller central energy densities with the decrease of $\beta$. From a numerical point of view it is very difficult to construct the whole branch if the   minimum ${\tilde \varepsilon_c}$ has very small value and that is why we have decided to perform the calculations for moderate $\beta$. 

The radius of the scalarized neutron stars plotted in Fig. \ref{fig:M_R} decreases compared to GR for small masses while it increases significantly for larger masses. This behavior is standard also for the scalarization in scalar-tensor theories with a single scalar field. Interestingly, the non-uniqueness of the scalarized solutions with respect to ${\tilde \varepsilon_c}$, that we discussed above, does not leave any significant imprint on the $M(R)$ dependence.

\begin{figure}
	\includegraphics[width=0.40\textwidth]{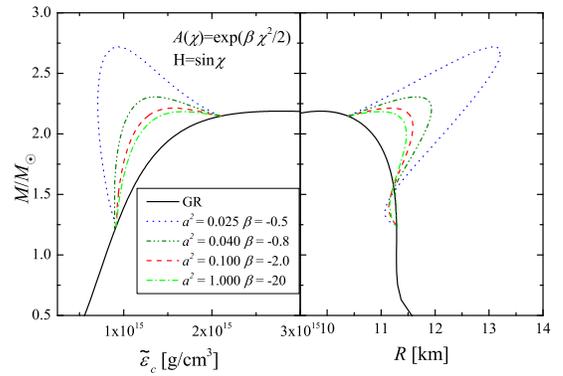}
	\caption{The mass as a function of the central energy density for the first branch of scalarized neutron stars, focusing on different combinations of $a^2$ and $\beta$, while keeping the ratio $\beta/a^2$ fixed.}
	\label{fig:M_rhoc_R_A1_varKappa}
\end{figure}

If $\beta$ is decreased further we have observed the appearance of a second branch of scalarized solutions characterized by scalar field with one node. For even smaller $\beta$ a third branch exists having scalar field with two nodes and so on.  The qualitative behavior of these branches is very similar to the nodeless scalar field branch. Such solutions possessing scalar field nodes are unstable, though, in other alternative theories where scalarization is observed \cite{Blazquez-Salcedo:2018jnn,Doneva2010}. Therefore, it is expected that they are unstable for the considered class of TMST as well and that is why  we will not present them explicitly.

In Fig. \ref{fig:M_rhoc_R_A1_varKappa} we have explored how a change of $a^2$ influences the results for a fixed form of  $A(\chi)$ and $H(\chi)$ having the following motivation in mind. Even though the perturbation equation \eqref{eq:PertEq} and Fig. \ref{fig:Bifurcations} depend only on the ratio ${\tilde \beta}/a^2$, clearly this is no longer true if we consider the full system of reduced field equation. Thus, even though the position of the bifurcation points depend only on  ${ \beta}/a^2$, the properties of the full scalarized branch of neutron stars is very sensitive to the particular values of ${ \beta}$ and $a^2$. In Fig. \ref{fig:M_rhoc_R_A1_varKappa} we have decided to keep the ratio ${ \beta}/a^2$ constant in order to preserve the position of the bifurcation points. As one can see the qualitative behavior of the branch changes considerably and while for large enough $a^2$ we have standard behavior of the scalarized branch and the neutron stars are uniquely determined by their central energy density, for smaller $a^2$ non-uniqueness of the scalarized solutions with respect to ${\tilde \varepsilon_c}$ is observed. 

The study of the parameter space presented above is by no means exhaustive. We have decided to show only some representative examples, though, because this paper is focused more on the existence of such solutions and their basic properties. The construction of such scalarized neutron star branches is very difficult and time consuming. That is why further studies should be done in parallel with study of the linear stability of the scalarized neutron stars. In this way one can concentrate on the physically relevant stable solution. Such a study is underway.

\begin{figure}
	\includegraphics[width=0.40\textwidth]{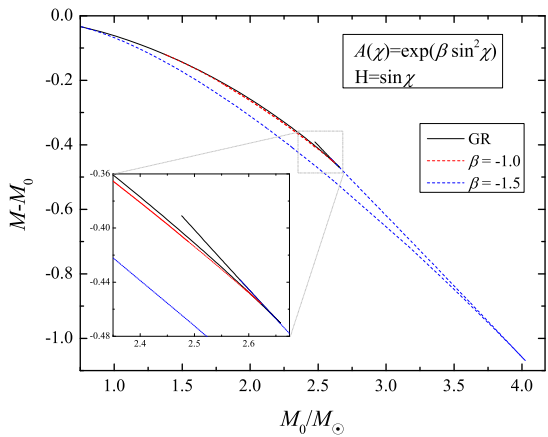}
	\caption{The binding energy as a function of the neutron star rest mass for some of the branches depicted in Fig. \ref{fig:M_rhoc}.}
	\label{fig:M0_M}
\end{figure}

A good indicator for the stability of the solutions is  the  biding energy of the neutron stars. Naturally, the solutions with larger absolute value of the binding energy will be  more stable and a change of stability is observed at the point were a cusp appears in the diagram relating the binding energy $M-M_0$ to the neutron star rest mass $M_0$. Such plot is presented in Fig. \ref{fig:M0_M}  for some representative branches and the picture is qualitatively similar for the rest of the cases we have studied. As one can see the scalarized branches have always larger absolute value of the binding energy compared to pure GR and therefore they should be the ones that realize in practice. This an expected result that holds for most of the other theories admitting scalarization \cite{Damour2013,Doneva:2017duq}. More interestingly, the presence of nonuniqueness within a single  scalarized branch of solutions for small $\beta$ is not manifested in Fig. \ref{fig:M0_M} and the corresponding branches look very similar to the large $\beta$ branches. Therefore, one can not conclude solely on the basis of the binging energy which one of the three solutions (the GR one and the two scalarized ones)  would realize in practice. In addition, for all of the scalarized branches only one cusp is observed that corresponds to the turning point in the $M({\tilde \varepsilon}_c)$ diagram. Therefore, similar to the GR solutions, a change of stability should be observed after the maximum of the mass is reached. The binding energy plot can show us only a signal of instability at a turning point and of course it can not be used to rigorously prove stability. Since we have non-uniqueness of the scalarized branches for certain ranges of the parameters it is very important to study the linear perturbations of the solutions that will be done in a future publication.

\begin{figure}
	\includegraphics[width=0.48\textwidth]{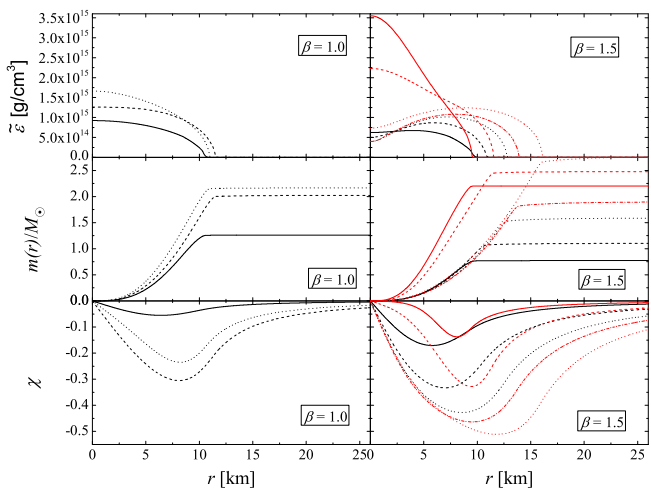}
	\caption{The radial profiles of the central energy density, the local mass and the scalar field for some representative solutions belonging to the branches depicted in the bottom panel of Fig. \ref{fig:M_rhoc},  for $A(\chi)=e^{\beta \sin^2{\chi}}$, $H=\sin{\chi}$ and $a^2=0.1$. In the \textit{left panel} the case of $\beta=1.0$ is presented while in the \textit{right panel} -- $\beta=1.5$ for which we have non-uniqueness of the scalarized solutions belonging to one single branch. In the right panel we have depicted with black (red) lines solutions belonging to the part of the branch before (after) the minimum of the central energy density observed in Fig. \ref{fig:M_rhoc}.}
	\label{fig:Solutions_A6_H0}
\end{figure}

The radial profiles of the energy density, the local mass defined as $m(r)=r (1-e^{-2\Lambda})/2$, and the scalar field $\chi$ for some representative solutions for $A(\chi)=e^{\beta \sin^2{\chi}}$, $H=\sin{\chi}$, $a^2=0.1$ and two values of $\beta$ are plotted in Fig. \ref{fig:Solutions_A6_H0}. In the left panels the $\beta=0.1$ case is depicted while in the right panels we plot the $\beta=0.15$ case where non-uniqueness of the scalarized neutron stars within a single branch of solutions in observed. In the former case the solution do not posses any peculiarities while for $\beta=0.15$  one can observe the appearance of an energy density maximum in the interior of some models. Thus the question whether the corresponding neutron stars are stable or not should be answer only after studying their radial stability.

At the end, let us make a comparison between the topological neutron stars \cite{Doneva:2019ltb} and the scalarized solutions obtained in the present paper. Technically, they originate from the same class of TMST of gravity in the particular case when the target space is $\mathbb{S}^3$ and different boundary conditions are applied. In the former case $n>0$ and we have topologically nontrivial solutions while in the latter case $n=0$ and the neutron stars are topologically trivial. Naturally, the pure general relativistic neutron stars are solutions of the field equations only for  $n=0$ that allows for scalarization. The structure of solutions with nontrivial scalar field also differs a lot -- in the topologically nontrivial case we can have a large number of branches but at least for the  cases studied in \cite{Doneva:2019ltb} the scalar field has no nodes. This is in contrast to the  $n=0$ case where more than one bifurcation from the general relativistic solutions can be observed  and the different branches of solutions can be labeled by the number of scalar field nodes. 

\section{Conclusions}

In the present paper we studied neutron stars in a particular class of tensor-multi-scalar-theories which admit scalarization. More specifically, we focused on the case when the target space manifold is a 3-dimensional symmetric space, namely $\mathbb{S}^3$, $\mathbb{H}^3$ or $\mathbb{R}^3$, motivated by the fact that these are among the simplest target spaces admitting spherically symmetric neutron star solutions
for the nontrivial map $\varphi : \text{\it spacetime} \to \text{\it target space}$ considered by us. The boundary conditions we impose are that the solutions are regular at the origin and asymptotically flat at infinity. In the particular case of $\mathbb{S}^3$ target space these requirements lead to a more general scalar field boundary condition of the type $\chi(0)=n\pi$ with $n\in \mathbb{Z}$. While for $n>0$ we have topologically nontrivial solutions\cite{Doneva:2019ltb}, in the present paper we concentrate on $n=0$ in which case the field equations always admit the GR solution. For certain ranges of the parameter space the pure Einstein neutron stars loose their stability and scalarized solutions appear. In order to be able to determine the central energy densities where the stability is lost we have solved numerically the equation governing the scalar field perturbations. This equation decouples from the rest of the perturbation equations, that are in fact the same as in GR.

We could obtain unstable modes of the GR solutions appearing at certain critical central energy densities which are exactly the points where new branches of scalarized solutions bifurcate form the GR ones. Naturally, more than one unstable mode can exist and the number of nodes of the corresponding scalar field perturbation is equal to the number of scalar field nodes of the corresponding scalarized solutions. The results show that a development of nontrivial scalar field happens both for positive and negative values of the parameter in the conformal factor $\beta$. While for negative $\beta$ we could find two bifurcation points for every branch -- one at smaller central energy densities, where the branch appears, and one at larger ${\tilde \varepsilon_c} $, where it merger again with the GR solutions, for positive $\beta$ only one such bifurcation point exists at least within a reasonable range of central energy densities. Even more interesting, for certain ranges of the parameters we can have not only several scalarized neutron stars with different scalar field nodes, but we also  nonuniqueness of the solutions belonging to the same scalarized branch. For example, for fixed parameters of the TMST and fixed ${\tilde \varepsilon_c}$ we can have two scalarized neutron stars possessing nodeless scalar field and one GR solutions. Even more interestingly, for such central energy densities the GR neutron stars are stable, even though the scalarized solutions alway have larger absolute value of the binging energy. This is quite different from the standard scalar-tensor theories with one scalar field where the scalarization leads to no more than one stable neutron star solution (either non-scalarized or scalarized one). All this  shows that the spectrum of solutions in TMST is very rich and complicated, and needs further investigation performed simultaneously with a linear stability analysis. Such a project underway.

Similar to other alternative theories of gravity admitting scalarization, the neutron stars with nontrivial scalar field are energetically move favorable over the GR ones. The above mentioned nonuniqueness is not manifested, though, in the behavior of the binding energy and thus it can not serve as a tool to discriminate between the different scalarized solutions.

To conclude let us briefly comment on the astrophysical implications. A very interesting and important property of the solutions is that the scalar field drops like $1/r^2$. Thus the scalar charge is zero and the binary pulsar experiments can not impose strong constraints on the parameters of the theory. This means that we can have much larger deviations from GR compared to the case of scalarized neutron stars in the standard scalar-tensor theories with one scalar field. This will certainly leave strong imprint on various observational properties of the scalarized TMST neutron stars and specially on the gravitational wave emission due to the possibility to have breathing modes in these theories of gravity. 

\section*{Acknowledgements}
DD acknowledges financial support via an Emmy Noether Research Group funded by the German Research Foundation (DFG) under grant
no. DO 1771/1-1. DD is indebted to the Baden-Wurttemberg Stiftung for the financial support of this research project by the Eliteprogramme for Postdocs.  SY would like to thank the University of Tuebingen for the financial support.  
The partial support by the Bulgarian NSF Grant DCOST 01/6 and the  Networking support by the COST Actions  CA16104 and CA16214 are also gratefully acknowledged.


\end{document}